\documentclass[sigconf,nonacm]{acmart}
\AtBeginDocument{%
  }
    
\usepackage{pifont}
\usepackage{url}
\usepackage{subfigure}
\usepackage{caption}
\usepackage{bigstrut,bigdelim}
\usepackage{float}  
\usepackage[figuresright]{rotating}
\usepackage{booktabs}
\usepackage{algorithmic}
\usepackage{algorithm}

\usepackage{wasysym}
\DeclareMathOperator*{\argmax}{arg\,max}

\usepackage[most]{tcolorbox}
\newtcbox{\mybox}[1][red]
  {on line, arc = 0pt, outer arc = 0pt,
    colback = #1!10!white, colframe = #1!50!black,
    boxsep = 0pt, left = 1pt, right = 1pt, top = 2pt, bottom = 2pt,
    boxrule = 0pt, bottomrule = 1pt, toprule = 1pt}
\definecolor{BoxBackground}{RGB}{240, 240, 240} 
\definecolor{BoxFrame}{RGB}{0, 0, 0} 
\definecolor{TitleBackground}{RGB}{0, 0, 0} 
\definecolor{TitleText}{RGB}{255, 255, 255} 
\tcbset{
  academicbox/.style={
    boxsep=5pt,
    left=2pt,
    right=2pt,
    bottom=0.5pt,
    boxrule=0.5pt,
    colback=BoxBackground,
    colframe=BoxFrame,
    colbacktitle=TitleBackground,
    coltitle=TitleText,
    enhanced,
    attach boxed title to top left={yshift=-0.1in,xshift=0.1in},
    boxed title style={boxrule=0pt,colframe=white},
    title={#1},
  }
}
\newtcolorbox{AcademicBox}[1][]{academicbox=#1}

\setcopyright{acmlicensed}
\copyrightyear{2018}
\acmYear{2018}
\acmDOI{XXXXXXX.XXXXXXX}
\acmConference[Conference acronym 'XX]{Make sure to enter the correct
  conference title from your rights confirmation email}{June 03--05,
  2018}{Woodstock, NY}
\acmISBN{978-1-4503-XXXX-X/2018/06}

\begin{document}

\title{Uncertainty-Guided Chain-of-Thought for Code Generation with LLMs}

\author{Yuqi Zhu}
\affiliation{%
  \institution{Peking University}
  \state{Beijing }
  \country{China}}

\author{Ge Li *}
\affiliation{%
  \institution{Peking University}
  \state{Beijing }
  \country{China}}

\author{Xue Jiang}
\affiliation{%
  \institution{Peking University}
  \state{Beijing }
  \country{China}}

\author{Jia Li}
\affiliation{%
  \institution{Peking University}
  \state{Beijing }
  \country{China}}

\author{Hong Mei }
\affiliation{%
 \institution{Peking University}
  \state{Beijing }
  \country{China}}

\author{Zhi Jin}
\affiliation{%
  \institution{Peking University}
  \state{Beijing }
  \country{China}}

\author{Yihong Dong}
\affiliation{%
  \institution{Peking University}
  \state{Beijing }
  \country{China}}

\renewcommand{\shortauthors}{Trovato et al.}

\begin{abstract}

Chain-of-Thought (CoT) reasoning has been demonstrated as an effective technique for improving the problem-solving capabilities of large language models (LLMs) in the context of code generation. However, existing CoT methods often exhibit a tendency toward "overthinking", where the LLM consistently applies reasoning strategies without adequately considering the task's underlying complexity. This results in the LLMs allocating excessive computational resources, in terms of tokens, to relatively simple tasks or problems where the correct answer is already evident. Additionally, this overthinking may lead LLMs down incorrect reasoning paths, resulting in incorrect code generation.
In this paper, we introduce UnCertainty-Aware Chain-of-Thought (UnCert-CoT), an LLM-based approach designed to enhance code generation by incorporating an uncertainty-aware CoT reasoning mechanism, which focuses computational resources on targeting points where LLMs are more prone to error.
We propose two confidence-based uncertainty measures: Entropy-based and Probability Differential-based methods. 
When uncertainty is high, UnCert-CoT activates CoT-decoding to generate multiple reasoning paths and selects the final code that exhibits the highest likelihood of correctness. 
In contrast, LLM directly generates the code when uncertainty is low.
This uncertainty judgment mechanism allows LLMs to prioritize complex tasks and avoid unnecessary steps in simpler cases, thereby improving overall efficiency and accuracy in code generation.
Our experimental results demonstrate that UnCert-CoT significantly enhances code generation accuracy on challenging benchmark MHPP(Mostly Hard Python Problems), it achieves improvements up to 6.1\% on PassRate accuracy, particularly in situations where traditional LLMs are prone to errors.
\end{abstract}



\maketitle

\section{Introduction}

Recent advances in Chain-of-Thought (CoT) reasoning have revolutionized the approach of large language models (LLMs) to tackle complex reasoning tasks. By decomposing problem-solving into sequential intermediate steps, CoT enables LLMs to emulate human-like logical deduction processes, significantly enhancing performance in problem-solving \cite{ZhouMHPPCB23} and reasoning \cite{cot}. The success of LLMs like DeepSeek-R1 \cite{deepseek} has demonstrated CoT's potential to address sophisticated reasoning challenges. The application of CoT has recently extended to the domain of code generation, with methods like Structured-CoT \cite{li2025structured} and Self-Planning \cite{selfplanning} offering innovative ways to enhance reasoning in code generation.
\begin{figure*}[tbp]
    \centering
    \includegraphics[width=\linewidth]{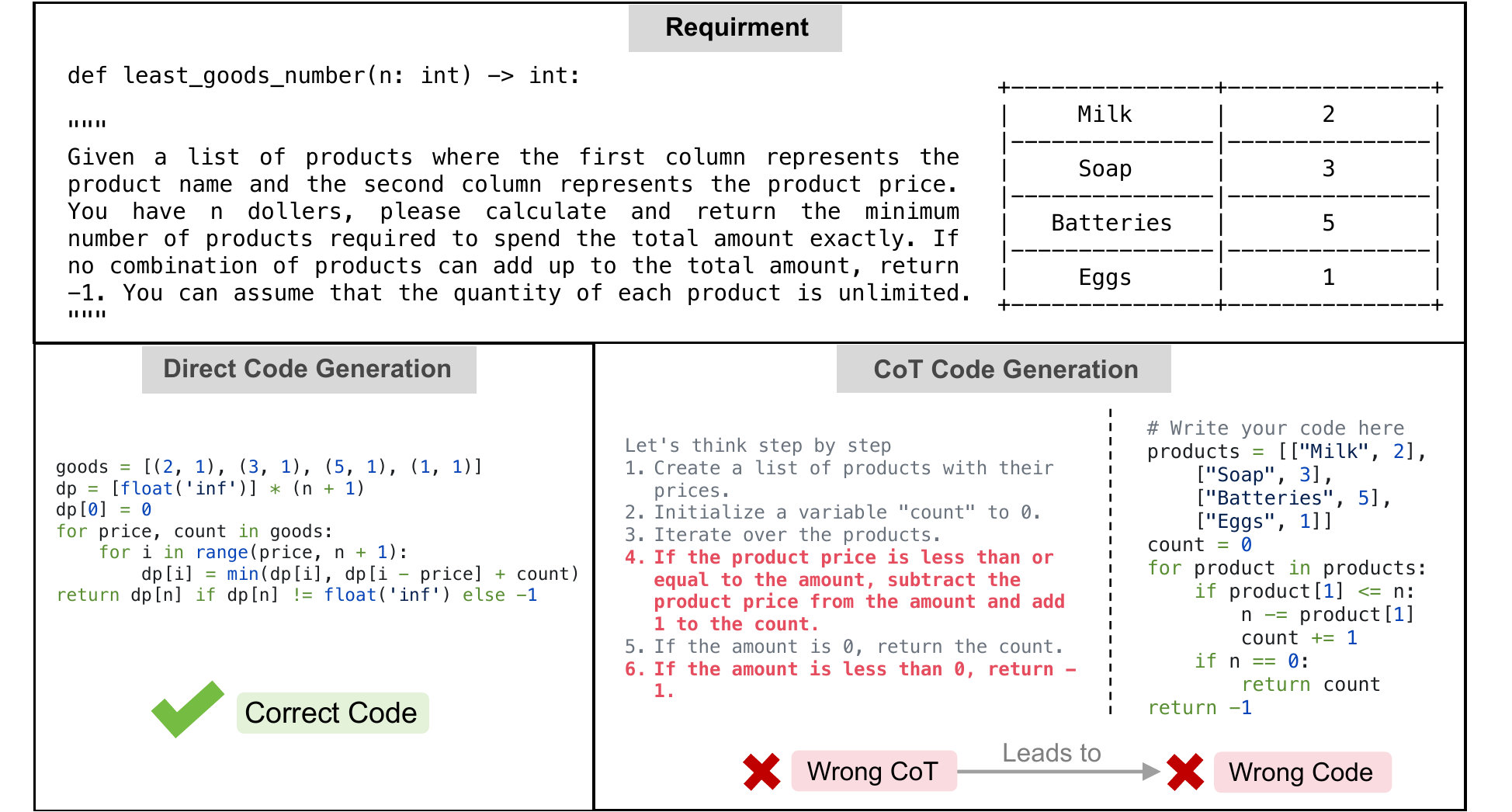}
    \caption{An illustration of the overthinking phenomenon of CoT Code Generation methods. The LLM could have answered a question correctly, however, utilizing CoT Code Generation methods makes the LLM produce an incorrect reasoning chain and generate incorrect code answers.}
     \label{fig:motivating example}
\end{figure*}

A key challenge of current CoT methods lies in the \textit{overthinking} phenomenon identified in the work \cite{overthink}. Overthinking occurs when LLMs, despite having sufficient knowledge to generate correct answers directly, produce unnecessary reasoning steps. 
These unnecessary reasoning steps may distort the reasoning process and paradoxically raise the error rate.
As demonstrated in Figure \ref{fig:motivating example}, a LLM that could have answered a question correctly ends up producing an incorrect result after applying the CoT method, due to the generation of an erroneous reasoning chain. Moreover, this issue results in inefficient allocation of computational resources during the generation process. Resources should be allocated to areas that require deeper reasoning, while the counterparts, that can directly obtain the correct results, should not consume additional resources.

To address these challenges, we introduce UnCertainty-Aware Chain-of-Thought (UnCert-CoT), a decoding method that dynamically invokes the CoT reasoning process based on the model's uncertainty during generation. The key challenge lies in identifying difficult points within the LLM’s generation process and selectively applying CoT to guide the model through these complexities. Prior research \cite{adapt} has highlighted that the prediction difficulty is particularly high at the beginning of a new line of code, as this position determines the subsequent logical structure the LLM construct next. Effectively handling this step places substantial demands on the LLM’s reasoning capabilities.
Therefore, we conduct uncertainty detection at the critical juncture where the LLM generates a new line of code. When the LLM’s uncertainty exceeds a pre-defined threshold, we invoke a CoT-decoding reasoning method which can guide the LLM to generate code that exhibits a relatively high likelihood of correctness.

To quantify uncertainty, we introduce two uncertainty measures: Entropy-based method and Probability Differential-based method. The Entropy-based approach estimates uncertainty by computing the information entropy \cite{entropy} of LLM’s prediction for the first non-indentation token in each new line of code. The Probability Differential-based method quantifies uncertainty based on the difference between the highest and second-highest predicted probabilities. These measures enable the LLM to identify situations where the LLM has high uncertainty levels, thereby triggering the reasoning process.

In cases of high uncertainty, UnCerT-CoT utilizes a CoT-decoding method to generate the subsequent line of code through a reasoning process. The CoT-decoding approach generates multiple reasoning steps, guiding the LLM to produce several potential solutions. Previous work \cite{codecalibration} has demonstrated a calibration between the confidence value and the correctness of the generated code. 
Therefore, we select the result with high confidence values among the reasoning results. On the other hand, when uncertainty is low, the LLM proceeds directly with code generation, thus ensuring efficiency and avoiding overthinking.

The UnCerT-CoT offers three key advantages: 1) Targeted reasoning: By dynamically invoking the reasoning method, UnCerT-CoT alleviates errors stemming from the LLM's difficulty in accurately predicting the first token of a code line. 2) Efficiency preservation: UnCerT-CoT allows the LLM to proceed directly with code generation when uncertainty is low, thereby maintaining efficiency. 3) Error mitigation: UnCerT-CoT addresses the issue of error accumulation in extended reasoning chains. For more complex tasks, it decomposes the reasoning process into line-step-level reasoning, generating multiple reasoning paths and selecting the most confident outcome at each stage, ensuring that each reasoning step is optimized and reducing the risk of error propagation.

We evaluated the performance of UnCerT-CoT and baseline approaches on two datasets: HumanEval \cite{codex}, and MHPP \cite{MHPP}. Our method demonstrates significant improvements, outperforming the baseline by 3.5\% and 6.1\% on the HumanEval and MHPP benchmarks, respectively. Additionally, we tested UnCerT-CoT on LLMs of varying sizes across different series, including DeepSeek-Coder \cite{DeepSeek-Coder}, CodeLlama \cite{codellama}, and Qwen-Coder \cite{Qwen-Coder}, observing consistent performance improvements. Extensive experiments confirm that UnCerT-CoT is robust to changes in hyperparameter configurations. Additionally, we evaluated the effectiveness of the proposed CoT-decoding method.

In summary, this paper makes the following contributions:
\begin{itemize}
    \item We identify the issue of overthinking in existing Chain-of-Thought (CoT) methods for code generation.  
    \item We propose a novel code generation approach called UnCert-CoT, which strategically incorporates uncertainty quantification during the code generation process to identify challenging aspects and selectively activates CoT to scale the reasoning process based on the identified challenges.
    \item We introduce two uncertainty computation approaches, including Entropy-based and Probability Differential-based methods, to detect high-uncertainty scenarios and trigger reasoning dynamically. 
    \item We demonstrate the consistent improvement of our method on two code generation datasets, using code LLMs of various sizes and from different families.
\end{itemize}

\section{Motivating Example}\label{motivating_example}

LLMs have demonstrated remarkable capabilities in code generation tasks. However, our investigation reveals a counterintuitive phenomenon: the inclusion of intermediate reasoning steps—often considered beneficial for complex problem-solving—can sometimes lead to degraded performance in code generation tasks. We term this the "overthinking phenomenon," where explicit step-by-step reasoning introduces errors that would not occur with direct code generation.

Figure \ref{fig:motivating example} illustrates this phenomenon through a representative example involving a dynamic programming problem. The task requires calculating the minimum number of products needed to spend exactly n dollars, given a set of products with fixed prices. When the LLM directly generates code for this requirement, it correctly implements a dynamic programming solution that exhaustively explores all possible product combinations to find the optimal answer. However, when prompted to first generate a CoT reasoning process before producing code, the model introduces a critical flaw. In step 4 of its reasoning chain, the LLM incorrectly adopts a greedy algorithm approach, attempting to subtract product prices from the total amount and incrementing a counter. This greedy strategy fails to consider alternative product combinations that might yield a more optimal solution. The reasoning error propagates to step 6, where the model prematurely concludes that a negative remaining amount indicates impossibility, rather than recognizing the need to explore alternative combinations.

This example highlights a fundamental limitation in current CoT approaches for code generation tasks. Our findings suggest the need for more nuanced prompting strategies that can selectively apply reasoning based on problem characteristics, rather than assuming CoT will universally enhance performance.
\section{Methodology}
\begin{figure*}[tbp]
    \centering
    \includegraphics[width=0.87\linewidth]{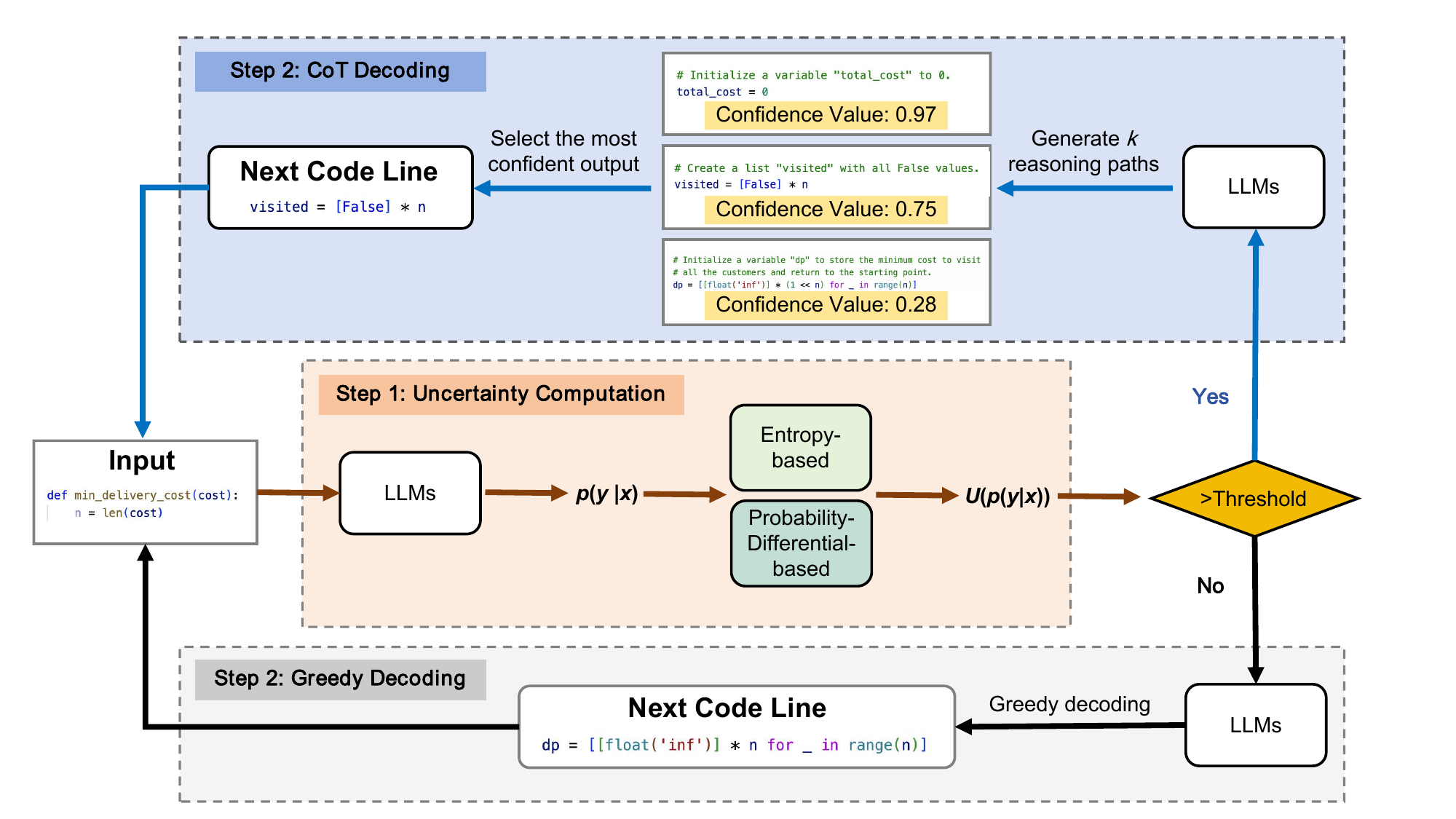}
    \caption{Overview of UnCert-CoT. When generating the next code line, UnCert-CoT first calculates the uncertainty value, if the uncertainty value exceeds a predefined threshold, it suggests that the LLM is uncertain about the next step. Consequently, the CoT-decoding method is triggered to provide additional reasoning steps, thereby enhancing the accuracy of the final output.}
    \label{fig:overview}
\end{figure*}
In this section, we delve into the details of our proposed Uncert-CoT method. We begin by providing an overview of the method and its overall workflow. Subsequently, we elaborate on the design principles of the two main components of our method: Uncertainty Computation and CoT-Decoding.

\subsection{Overview}
Our proposed method aims to enhance the accuracy of code generation by selectively invoking the CoT method during the code generation process of LLMs. 
Previous work \cite{adapt} has highlighted that the prediction difficulty at the beginning of a new line of code is notably higher than at other positions, as it determines the subsequent logical structure that the LLM needs to implement. This task places a significant demand on the LLM's reasoning capabilities.
Therefore, we perform uncertainty detection at the critical juncture where the LLM generates a new line of code. When the LLM’s uncertainty exceeds a pre-defined threshold, we dynamically invoke a CoT-decoding reasoning method which can guide the LLM to generate code that exhibits a relatively high likelihood of correctness. The overview our our method is shown in Figure \ref{fig:overview}.

The methodology consists of two main components. The first component is the \textbf{uncertainty computation} method, which determines when to invoke the CoT method during the generation process. We propose two uncertainty measures: the Entropy-based method and the Probability Differential-based method. The entropy-based method evaluates the uncertainty by measuring the randomness of the LLM's output distribution, the probability differential-based method assesses the difference in probabilities between the most likely and the second most likely tokens to determine the level of uncertainty.

The second component is the \textbf{CoT-decoding} method. This part employs the CoT-decoding method to generate multiple potential solutions for the next line of code through reasoning. Among these, the solution with the highest confidence score is selected as the final output. This ensures that the LLM leverages the most reliable CoT to improve the overall accuracy and reliability of the generated code.

\subsection{Uncertainty Computation}

In this section, we present our methodology for calculating uncertainty value $U(p)$, drawing inspiration from existing literature on uncertainty estimation in LLMs \cite{confidence}. 
We have developed two methods for uncertainty measurement: the Entropy-Based method and the Probability Difference-Based method. In the following, we provide a detailed explanation of each approach.

\subsubsection{Entropy-based uncertainty measurement}
In information theory, the entropy of a random variable quantifies the average level of uncertainty associated with the variable's potential states or possible outcomes\cite{entropy}. In the context of machine learning, the LLM typically outputs a probability distribution over possible outcomes, representing its belief about the likelihood of each outcome. Therefore, we can use entropy to measure the degree of uncertainty of a probability distribution predicted by LLMs. 

The calculation of entropy $ H(p) $ for a probability distribution $ p = \{p_1, p_2, ..., p_n\} $ is defined as:
\begin{equation}
    H(p) = - \sum_{i=1}^{n} p_i \log(p_i)
\end{equation}
 $p_i $ represents the probability assigned to the  $i$ -th token of the LLM's vocabulary.
 
When the LLM is uncertain about its predictions, the probabilities for all outcomes are evenly distributed, which results in a high entropy value. For instance, if the LLM predicts multiple possible outcomes with equal probabilities, it results in the highest entropy value $\log N$ ($N$ is the possible value number of the random variable).


Based on the theoretical foundation laid out previously, we employ entropy to define our uncertainty computation method $U_e(p)$. Specifically, we quantify the uncertainty of the LLM's predictions using the entropy of the probability distribution over possible tokens.
Given an input $x$ and the context $y_1,y_2,…,y_{n-1}$ that has already been generated, we consider the prediction probability $p$ of the $n$-th token generated by the LLM. The uncertainty value $U_e(p)$ at this step is computed using the following formula:

\begin{equation}
    U_e(p)=\frac{H_n(p)}{\log V}
\end{equation}
\begin{equation}
    H_n(p) = -\sum_{i=1}^{V}p(y_n^i|x,y_1,…,y_{n-1})\log p(y_n^i|x,y_1,…,y_{n-1})
\end{equation}
$p(y_n^i|x,y_1,y_2,…,y_{n-1})$ represents the conditional probability of LLM generating the $n$-th token $y_n$ given the input $x$ and the previously generated context $y_1,y_2,…,y_{n-1}$. The summation is over all possible tokens in the vocabulary. $V$ is the vocabulary size of the LLM. 

The uncertainty computation function $U_e(p)$ that we have defined possesses several key properties that are crucial for its effective application in our framework:
1) The function $U_e(p)$ is normalized to $[0,1]$. This normalization ensures that the uncertainty values are consistently interpretable across different contexts and scales.
2) When the LLM's uncertainty is at its maximum, i.e. $p=(\frac{1}{V}, \frac{1}{V},.., \frac{1}{V})$, the entropy value of $H(p)$ is equals to $\log V$, then $U_e(p)=1$. Suggesting that the LLM is highly uncertain about the next token to generate.
Conversely, when the LLM's uncertainty is at its minimum, i.e. $p=(1.0, 0.0, ...0.0)$, the value of entropy $U_e(p) = H(p)=0$. 

By defining $U_e(p)$ in this manner, we ensure that the LLM's uncertainty is uniformly distributed across the range [0, 1]. This uniform distribution facilitates a more intuitive and consistent interpretation of the uncertainty values. Specifically, it allows us to set clear thresholds for invoking the CoT method based on the normalized uncertainty values. For instance, if $U_e(p)$ exceeds a certain threshold (e.g., 0.5), it indicates a significant level of uncertainty, suggesting the use of the CoT method to enhance the accuracy of the generated output.

\subsubsection{Probability Differential-based uncertainty measurement}

In this section, we utilize an uncertainty measurement $U_d(p)$ based on the difference in predicted probabilities following the method \cite{cotdecoding}. Specifically, we calculate the difference between the highest predicted probability $y_n^1$ and the second-highest predicted probability $y_n^2$ in the LLM's prediction for a specific position, using this difference as the uncertainty indicator for that position. The equation of $U_d(p)$ is defined as followed:
\begin{equation}
    U_d(p) = 1-(p(y_n^1|x,y_1,…,y_{n-1}) - p(y_n^2|x,y_1,…,y_{n-1}))
\end{equation}
A larger $U_d(p)$ value indicates that the LLM has greater certainty at the current step, suggesting that the LLM is less uncertain about its current choice. In contrast, a smaller difference indicates that the LLM has to greater uncertainty in its current choice.

The value of $U_d(p)$ ranges from 0 to 1, when LLMs are most certain about the output answer, i.e. $ p = (1.0, 0.0, ..., 0.0)$, the  $U_d(p) = 0$. When LLMs is uncertain about the answer, i.e. $p=(\frac{1}{V}, \frac{1}{V},.., \frac{1}{V})$, $U_d(p) = 1$. 

\subsection{CoT-decoding}


In this section, we present the CoT-decoding method used in UnCert-CoT. 
Our CoT-decoding method aims to improve the accuracy of LLM-generated results when the LLM exhibits high uncertainty in its current generation. Specifically, the method improves the confidence score of the LLM's output by exploring multiple reasoning paths. 
The CoT-decoding method we propose is an adaptation of the decoding method \cite{cotdecoding}, which leverages the LLM's intrinsic reasoning capabilities to generate solutions and selects the final output based on the confidence scores associated with each generated result. The authors conduct extensive experiments and observe that selecting answers with higher confidence scores tends to yield higher accuracy rates. This phenomenon has also been identified in the context of code generation tasks \cite{codecalibration}.

However, the original CoT-Decoding method is not directly applicable to code generation tasks. 
This is because the CoT-Decoding method was initially designed for natural language generation, where the LLMs do not need few-shot prompting to generate reasoning paths. As for code generation tasks, the LLM's distribution tends to favor the continuation of code with more code. To address this challenge, we utilize the in-context learning capability\cite{icl} of the LLM and employ a few-shot prompting with $m$ demonstration examples to guide the LLM in sampling $k$ reasoning paths with temperature $t$ for the next line of code. Then, we calculate the confidence score for each reasoning path. The result with the highest confidence score is chosen as the final output. 

The method can be formulated as follows:
\begin{equation}
    (r_i, c_i) \sim M( \{x_1', r_1', c_1'\}, \dots,\{x_m', r_m', c_m'\}, x),i = 1, 2, \dots, k
\end{equation}
\begin{equation}
    \hat{c} = \argmax Confidence(c_i),  \quad   i = 1, 2, \dots, k
\end{equation}
\begin{equation}
    Confidence(c_i) = \frac{1}{len(c_i)}\sum_{y_t\in c_i}(p(y_t^1|y_{t-1}, x)-p(y_t^2|y_{t-1}, x))
\end{equation}

Where $M$ is the LLM, $\{x_1', r_1', c_1'\}, \dots,\{x_m', r_m', c_m'\}$ are the few-shot demonstrate examples. $r_i$ is the $i$-th reasoning path generated by LLM, $c_i$ is the code generated corresponding to $r_i$.  $y_t$ is the $t$-th token in the code $c_i$.

\subsection{UnCert-CoT}
We introduce the overall pipeline of UnCert-CoT, which is shown in Figure \ref{fig:overview}. When generating code with LLM, before generating the next code line, we compute the current uncertainty of LLM about the next step $U(p)$. If $U(p)>\tau$ (a predefined threshold), we trigger the CoT-decoding process, which guides the LLM to generate reasoning paths first and then result in more confident next code lines.  In contrast, if $U(p)\leqslant \tau$, we use a greedy search method to generate the next code line, which selects the max probability token at each token step.

\section{Study Design}

To verify the effectiveness of UnCert-CoT, we conduct a large-scale study to answer four research questions. Next, we will describe the details of our study, including datasets, baselines, LLMs, and evaluation metrics.

\subsection{Research Questions}
Our study mainly answers the following research questions (RQ).

\textbf{RQ1: How does UnCert-CoT perform in code generation compared with the state-of-the-art CoT methods?}
RQ1 aims to evaluate the effectiveness of UnCert-CoT in comparison to state-of-the-art CoT methods.

\textbf{RQ2: How does UnCert-CoT perform in code generation with different LLMs?}
RQ2 evaluates the effectiveness of UnCert-CoT across a diverse range of LLMs varying in both size and architecture.

\textbf{RQ3: The ablation study of the CoT-decoding method.} RQ3 examines the impact of the CoT-decoding method in UnCert-CoT through an ablation study.

\textbf{RQ4: What is the performance of UnCert-CoT on different uncertainty threshold settings?} RQ4 assesses the robustness of UnCert-CoT across varying uncertainty threshold settings.

\subsection{Datasets} \label{datasets}
\begin{itemize}
    \item \textbf{HumanEval} \cite{codex} is a Python code generation benchmark. It consists of 164 original programming problems, assessing language comprehension, algorithms, and simple mathematics, with some comparable to simple software interview questions. Each problem consists of a manually written programming problem, which consists of a natural language requirement, a function signature, and several unit tests. 
    \item \textbf{MHPP (Mostly Hard Python Problems)} \cite{MHPP} consisting of 210 unique human-curated problems. MHPP gauges LLMs’ abilities to comprehend specifications and restrictions, engage in multi-step reasoning, and apply coding knowledge effectively by focusing on the combination of natural language and code reasoning. Additionally, the authors of MHPP developed an evaluation pipeline to mitigate data leakage. Initial assessments of 26 LLMs using MHPP revealed that many LLMs achieving strong performance on HumanEval struggled to attain comparable results on MHPP. Consequently, MHPP is a challenging benchmark for LLMs and exposes previously unrecognized limitations in various LLMs.
\end{itemize}

\subsection{Models} \label{models}

\begin{itemize}
\item  \textbf{DeepSeekCoder\cite{DeepSeek-Coder}} is composed of a series of code LLMs (ranging from 1B to 33B), each trained from scratch on 2T tokens, with a composition of 87\% code and 13\% natural language in both English and Chinese. Each LLM is pre-trained on project-level code corpus by employing a window size of 16K and an extra fill-in-the-blank task, to support project-level code completion and infilling. For coding capabilities, DeepSeek Coder achieves state-of-the-art performance among open-source code LLMs on multiple programming languages and various benchmarks. In this paper, we use the DeepSeekCoder-6.7B-Base model.

\item  \textbf{CodeLlama\cite{codellama}} is a family of LLMs for code based on Llama 2 providing state-of-the-art performance among open LLMs, infilling capabilities, support for large input contexts, and zero-shot instruction following ability for programming tasks. It provides multiple flavors to cover a wide range of applications: foundation models (CodeLlama), Python specializations (CodeLlama-Python), and instruction-following models (CodeLlama-Instruct) with 7B, 13B, and 34B parameters each. All LLMs are trained on sequences of 16k tokens and show improvements on inputs with up to 100k tokens. In this paper, we use the CodeLlama-13B-python-hf model.

\item  \textbf{QwenCoder\cite{Qwen-Coder}} is a new series of LLMs designed to achieve top-tier performance in coding tasks at various model sizes. Qwen2.5-Coder models are derived from the Qwen2.5 LLMs\cite{qwen}, inheriting their advanced architecture and tokenizer. These LLMs are trained on extensive datasets and further fine-tuned on carefully curated instruction datasets specifically designed for coding tasks. In this paper, we use the Qwen2.5-Coder-7B-Base model.
\end{itemize}

\subsection{Baselines} \label{baselines}
\begin{itemize}
\item  \textbf{Base Model \cite{cot}} uses the greedy search method to generate the final result.
\item  \textbf{Zero-shot CoT \cite{cot}} directly feeds the requirement into LLMs without examples. It adds a "\# Let's think step by step." line before generating the corresponding code. Then, it extracts a generated code from LLMs’ outputs.
\item  \textbf{Self-planning\cite{selfplanning}} provides a few demonstration examples to the LLM to assist the LLM in generating the correct code. 
\item  \textbf{CoT-Decoding\cite{cotdecoding}} uses the Zero-shot CoT to generate multiple samples with reasoning paths and select the answer with the highest confidence value as the final result. 
\end{itemize}

\subsection{Evaluation Metric} \label{metrics}

PassRate \cite{codex} measures the functional correctness of the generated code by executing test cases. For $N$ problems, if the LLM successfully generate correct answers for $C$ problems, PassRate is calculated as follows:
\begin{equation}
    \text{PassRate} = \frac{C}{N}
\end{equation}

\begin{table}[!t]
\setlength{\abovecaptionskip}{2mm} 
  \centering
  \caption{Comparison of the CoT methods' performances on code generation task. }
  \resizebox{0.99\linewidth}{!}{
    \begin{tabular}{rrcc}
    \toprule
          & \multicolumn{1}{l}{Methods} & HumanEval  & MHPP  \\
    \midrule
    \multicolumn{1}{l}{\multirow{4}{*}{Baselines}} & \multicolumn{1}{l}{Base Model~\cite{deepseek}} & 0.481   & 0.233    \\
          & \multicolumn{1}{l}{zero-shot CoT~\cite{cot}} & 0.408  &  0.162 \\
           &  \multicolumn{1}{l}{CoT-Decoding~\cite{cotdecoding}} &   0.335   &   0.148  \\
             & \multicolumn{1}{l}{Self-planning ~\cite{selfplanning}} & 0.542  & 0.190 \\
    \midrule
    \multicolumn{1}{l}{\multirow{4}{*}{UnCert-CoT}}
     & \multicolumn{1}{l}{Entropy-based } &  0.555   &  0.262  \\    
     && ($\uparrow$ \textbf{2.4\%})   & ($\uparrow$ \textbf{6.1\%})    \\
    & \multicolumn{1}{l}{P-D based} & 0.561 &  0.257\\
    && ($\uparrow$ \textbf{3.5\%})    & ($\uparrow$ \textbf{4.0\%})    \\

    \bottomrule
    \end{tabular}%
   } 
  \label{rq1}%

\end{table}%

\section{RESULTS AND ANALYSES}
In our first research question, we evaluate the performance of UnCerT-CoT with respect to other CoT methods on three code generation benchmarks. 

\textbf{RQ1: How does UnCert-CoT perform in code generation compared with the state-of-the-art CoT methods?}

\textbf{Setup.}
We compare UnCert-CoT with two advanced CoT approaches used in code generation tasks (Section \ref{baselines}). Then, we use PassRate (Section \ref{metrics}) to measure the performance of different approaches on two code generation benchmarks (Section \ref{datasets}). 
We use the DeepSeekCoder-6.7B-Base for RQ1 experiments.
To conserve space, we abbreviate Probability Differential-based as P-D based in all experimental results.

The generation setting of our experiments is 2-shot prompting. The prompt setting is to follow the previous work \cite{selfplanning}.
The temperature value $T$ of the baseline CoT-Decoding and our UnCert-CoT is 0.4, and the sampling number $m$ is 5. 

\textbf{Results.} The results are shown in Table \ref{rq1}. The values in parentheses are relative improvements compared to the state-of-the-art baseline.

\textbf{Analyses.} 
\textbf{(1) UnCert-CoT achieves the best performance among all approaches.}
In comparison to the state-of-the-art baselines, UnCert-CoT delivers notable relative improvements of 3.5\% and 6.1\% on the HumanEval and MHPP datasets, respectively, based on the PassRate metric. On average, the Entropy-based method achieves a 4.25 \% improvement, and the Probability Differential-based method provides a more modest 3.75\% improvement. These improvements not only demonstrate the effectiveness of our approach but also underscore the robustness of UnCert-CoT across diverse code generation tasks. 
\textbf{(2) Applying original CoT-Decoding directly to code generation tasks results in a degradation of performance. }In the context of CoT-Decoding, while it has shown strong performance in mathematical problem-solving, its application to code generation has not yielded similarly favorable results. The underlying reason for this discrepancy lies in the confidence-based approach CoT-Decoding employs to select the optimal outcome. In mathematical tasks, the LLM typically relies on the confidence of the final token to evaluate the correctness of the solution. However, when applied to code generation, the final result consists of a long code sequence, making it difficult to assess the quality of the code based solely on the overall confidence of the entire sequence. This challenge arises because, unlike mathematical problems where a single final answer is generated, code generation involves a series of tokens whose cumulative confidence does not directly correspond to the accuracy or effectiveness of the generated code. To address this issue, our approach focuses on treating code generation as a sequence of discrete steps or reasoning, rather than relying on the overall confidence of the complete code snippet. By evaluating the quality of code at each reasoning step, our method provides a more granular and effective assessment, which alleviates the limitations observed in CoT-Decoding\cite{cotdecoding} for code generation.
\textbf{(3) The UnCert-CoT method achieves better improvement on the benchmark that requires complex code reasoning abilities.}
As discussed in Section \ref{datasets}, the UnCert-CoT method demonstrates a more substantial improvement on the MHPP benchmark, which involves more intricate reasoning and logic compared to other benchmarks. The results indicate that the self-planning method, in contrast, deteriorates the performance of the MHPP dataset. Specifically, questions that the LLM could have answered correctly without self-planning perform worse after its application. This decline can be attributed to a significant limitation of the self-planning approach: it struggles to identify a universal prompt that is effective across diverse requirement formats, especially when the requirements are not homogeneous. Our method, however, avoids this issue by performing reasoning at the granularity of individual lines. Each reasoning step generates reasoning thought as a single comment, allowing for more targeted and flexible processing. This design mitigates the rigid constraints imposed by a fixed prompt format, resulting in more effective handling of complex code generation tasks. Consequently, our method achieves notable improvements of 6.1\% and 4.0\% on the MHPP benchmark, demonstrating its ability to handle challenging code generation tasks that require sophisticated reasoning. 

\begin{tcolorbox} 
  \textbf{Answer to RQ1:} UnCert-CoT achieves the best performance among all benchmarks. Moreover, the UnCert-CoT method achieves better improvement on the benchmark that requires complex code reasoning abilities.
\end{tcolorbox}


\begin{table}[tb]
  \centering
  \caption{Evaluation results of UnCerT-CoT with different LLMs on HumanEval and MHPP benchmarks.}
  \resizebox{0.99\linewidth}{!}{
    \begin{tabular}{lccc}
    \toprule
    \textbf{Models}& Method & HumanEval  & MHPP    \\
    \midrule
    \multicolumn{1}{l}{\multirow{5}{*}{DeepSeek-Coder\cite{deepseek}}}
    & \multicolumn{1}{l}{Base Model}   &0.481  &0.247  \\
    & \multicolumn{1}{l}{Entropy based} & 0.555 &0.262  \\
    &  &($\uparrow$ \textbf{15.4\%})   & ($\uparrow$ \textbf{6.1\%})     \\
    & \multicolumn{1}{l}{P-D based}& 0.561&0.257\\
    & &($\uparrow$ \textbf{16.6\%})    & ($\uparrow$ \textbf{4.0\%})    \\
    \midrule
    \multicolumn{1}{l}{\multirow{5}{*}{Qwen-Coder\cite{Qwen-Coder}}}
    & \multicolumn{1}{l}{Base Model}  &0.591 & 0.329  \\
    & \multicolumn{1}{l}{Entropy-based}  &0.598  & 0.338   \\
    & & ($\uparrow$ \textbf{1.2\%})& ($\uparrow$ \textbf{2.8\%})    \\
    & \multicolumn{1}{l}{P-D based} &0.598 & 0.333\\
    &  &($\uparrow$ \textbf{1.2\%})& ($\uparrow$ \textbf{1.2\%})    \\
    \midrule
        \multicolumn{1}{l}{\multirow{5}{*}{CodeLlama\cite{codellama}}}
    & \multicolumn{1}{l}{Base Model}  &  0.426 &0.162  \\
    & \multicolumn{1}{l}{Entropy-based}  & 0.463 &  0.190 \\
    &&($\uparrow$ \textbf{8.7\%}) & ($\uparrow$ \textbf{17.3\%})    \\
    & \multicolumn{1}{l}{P-D based}& 0.476&0.181\\
    & & ($\uparrow$ \textbf{11.7\%}) & ($\uparrow$ \textbf{11.7\%})    \\
    \bottomrule
    \end{tabular}%
    }
  \label{rq2}%
\end{table}%
\textbf{RQ2: How does UnCert-CoT perform in code generation with different LLMs? }

\textbf{Setup.}
We compare UnCert-CoT with three state-of-the-art code LLMs (Section \ref{models}). 
We use the DeepSeekCoder-6.7B-Base, Qwen2.5-Coder-7B-Base, and CodeLlama-python-hf-13B for RQ2 experiments. 
These three LLMs belong to different families and vary in the number of parameters. Additionally, all LLMs from these families currently demonstrate state-of-the-art performance in code generation tasks.
The generation setting is the same as RQ1.

\textbf{Results.} The results on three datasets are shown in Table \ref{rq2}. The values in parentheses are relative improvements compared to the baseline model.

\textbf{Analyses.} 
\textbf{(1) UnCert-CoT has demonstrated significant performance improvements across various large LLMs in our experiments. }We utilized LLMs of different sizes from distinct families, including DeepSeekCoder, Qwen-Coder, and CodeLlama, with parameter sizes of 6.7B, 7B, and 13B, respectively. On the HumanEval benchmark, our method achieved average improvements of 16.0\%, 1.2\%, and 10.2\% compared to the base LLMs for DeepSeekCoder, Qwen-Coder, and CodeLlama, respectively. On the MBPP benchmark, the performance gains over the base LLMs were 5.1\%, 2.0\%, and 14.5\% for the same LLMs. These results highlight the substantial effectiveness of UnCert-CoT in enhancing LLM performance across a range of LLMs.
\textbf{(2) The performance improvements of UnCert-CoT vary across different LLMs.}
This variation can be attributed to the inherent differences in the base capabilities of the LLMs. For instance, the Qwen-Coder\cite{Qwen-Coder} series has already achieved performance levels comparable to ChatGPT\cite{ChatGPT} and Claude 3.5 Sonnet\cite{Claude} on several code generation datasets. As a result, the improvements observed in simpler datasets, such as HumanEval, are less pronounced for these LLMs. 

\begin{tcolorbox} 
  \textbf{Answer to RQ2:} UnCert-CoT, demonstrates improvements across models of varying sizes and from different families. 
\end{tcolorbox}

\textbf{RQ3: Ablation study of the CoT decoding method. }

\textbf{Setup.} 
We compare UnCert-CoT with CoT-decoding and without CoT-decoding on HumanEval and MHPP benchmarks. 
UnCert-CoT without CoT-decoding represents the method that uses greedy decoding to generate a reasoning path and derive the generated code.  
We use the DeepSeekCoder-6.7B-Base for RQ3 experiments. 

\textbf{Results.} The experimental results are shown in Table \ref{rq3}. The values in parentheses are relative decrease in performance compared to the baseline model.
\begin{table}[tb]
  \centering
  \caption{Evaluation results of ablation study of the CoT-decoding method.}
  \resizebox{0.99\linewidth}{!}{
    \begin{tabular}{lccc}
    \toprule
    \textbf{Approaches}& Method & HumanEval  & MHPP    \\
    \midrule
    \multicolumn{1}{l}{\multirow{2}{*}{UnCerT-CoT}}
    & \multicolumn{1}{l}{Entropy based} & 0.555 &0.262  \\
    & \multicolumn{1}{l}{P-D based}& 0.561&0.257\\
    \midrule
    \multicolumn{1}{l}{\multirow{4}{*}{UnCerT-CoT(w)}}
    & \multicolumn{1}{l}{Entropy based} & 0.530 &0.229  \\
    &  &($\downarrow$ \textbf{4.5\%})   & ($\downarrow$ \textbf{12.6\%})     \\
    & \multicolumn{1}{l}{P-D based}& 0.524&0.243\\
    & &($\downarrow$ \textbf{6.5\%})    & ($\downarrow$ \textbf{5.4\%})    \\

    \bottomrule
    \end{tabular}%
    }
  \label{rq3}%
\end{table}%

\textbf{Analyses.} 

\textbf{The exclusion of CoT decoding results in a decrease in performance, especially on difficult problems. }
Excluding CoT decoding and relying solely on greedy decoding to generate a single reasoning path leads to a significant decrease in the effectiveness of the proposed method. As observed in the results shown in Table \ref{rq3}, the absence of greedy decoding results in an average performance drop of 5.5\% and 9\% on the HumanEval and MHPP datasets, respectively. This decline is particularly pronounced on more challenging problems (MHPP). The reason for this deterioration lies in the inherent complexity of harder problems, where the LLM typically needs to generate multiple reasoning steps to arrive at the correct solution. Without CoT decoding, the LLM's ability to iteratively refine its reasoning through successive steps is hindered, thus limiting its ability to solve more complex tasks effectively. In contrast, CoT decoding facilitates the generation of multiple reasoning paths, which significantly enhances the LLM’s ability to tackle challenging problems.

\begin{tcolorbox}
  \textbf{Answer to RQ3}: The exclusion of CoT decoding results in a decrease in performance, especially on complex coding benchmarks.
\end{tcolorbox}

\textbf{RQ4: What is the performance of UnCert-CoT on different uncertainty threshold $\tau$ settings?}

\textbf{Setup.}

In this section, we evaluate the performance of the UnCert-CoT method under different uncertainty thresholds $\tau$. Our experiments were conducted with the HumanEval and MHPP datasets and the DeepSeek-6.7B-Base model. We performed analytical experiments using both entropy-based and probability differential-based methods.

\begin{figure}[tbp]
    \centering
    \includegraphics[width=0.965\linewidth]{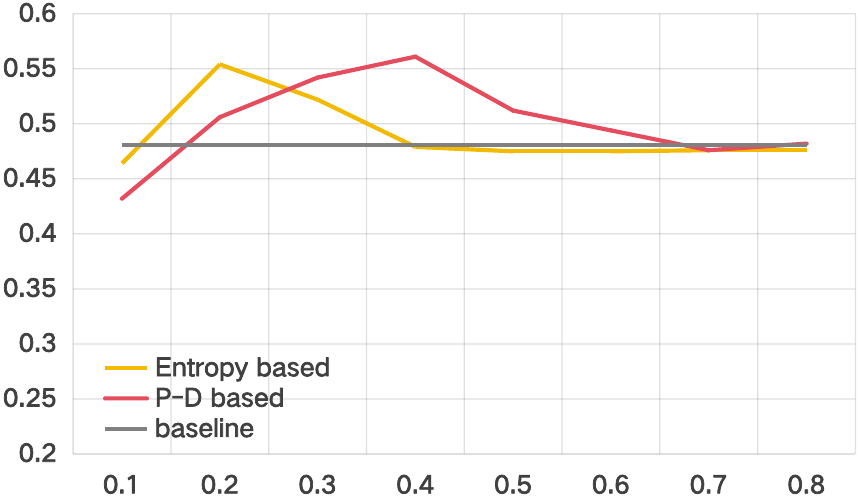}
    \caption{Performance of UnCert-CoT under different $\tau$ threshold settings on HumanEval dataset.}
    \label{hyper}
\end{figure}

\begin{figure}[tbp]
    \centering
    \includegraphics[width=0.965\linewidth]{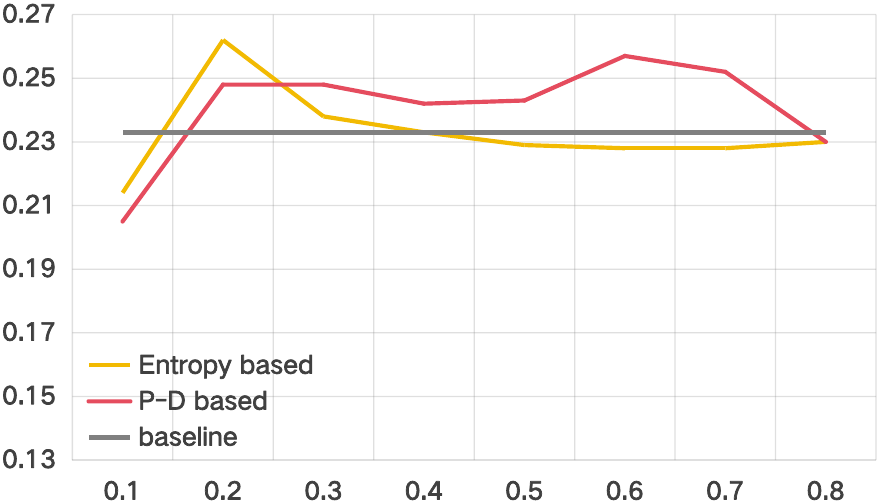}
    \caption{Performance of UnCert-CoT under different $\tau$ threshold settings on MHPP dataset.}
    \label{hyper_2}
\end{figure}

\textbf{Results.} The experimental results of hyperparameter analysis is in Figure \ref{hyper}, and Figure \ref{hyper_2}. The grey line represents the performance of the base LLM. The red line represents the performance of the Probability Differential-based UnCerT-CoT. The yellow line represents the performance of the Entropy-based UnCerT-CoT. The x-axis represents the values of the hyperparameter \(\tau\), while the y-axis denotes the pass rate metric.

\textbf{Analyses.} 
\textbf{(1) The performance of UnCert-CoT follows a predictable trend as the threshold $\tau$ is changed.} Initially, as the threshold $\tau$ increases, the performance of UnCert-CoT improves, reaching a peak at a certain point. However, beyond this threshold, the performance begins to decline. Specifically, when the threshold $\tau < 0.1$, the method’s performance is inferior to that of the baseline. This reduction can be attributed to the LLM's tendency to overthink during reasoning at each step. In scenarios where the solution is already clear, continuing the reasoning process beyond the point of clarity introduces unnecessary intermediate steps. These additional steps can result in the generation of flawed reasoning sequences, which in turn lead to the generation of incorrect outputs. The LLM’s over-reliance on excessive reasoning steps in such cases leads to a deterioration in its overall performance.
\textbf{(2) The UnCert-CoT method consistently outperforms the base model across multiple threshold settings.}
The UnCert-CoT method consistently outperforms the base model across a range of threshold $\tau$ settings. This consistent improvement highlights the robustness of our approach to variations in hyperparameter configurations, as it yields favorable results across different $\tau$ values. This suggests that the method’s performance is not overly dependent on fine-tuning a particular configuration.
\textbf{(3)The optimal threshold settings for the two UnCert-CoT methods exhibit differences. }Based on the experimental results, we observe that for the Entropy-based method, the improvements are achieved with threshold values $\tau \in [0.2, 0.3]$. On the other hand, the Probability Differential-based method yields improvements with threshold values $\tau \in [0.2, 0.7]$.
This divergence in optimal thresholds suggests that each method responds differently to variations in the threshold, with the Entropy-based method benefiting from a slightly smaller threshold range compared to the Probability Differential-based method. Consequently, these findings provide valuable insights for future work, as they offer a direction for further refinement in threshold optimization strategies. 

\begin{tcolorbox}
  \textbf{Answer to RQ4:}  The performance of UnCert-CoT follows a predictable trend as the threshold $\tau$ is changed. The UnCert-CoT method demonstrates a certain level of robustness to hyperparameter settings. Additionally, the optimal threshold range differs between the two uncertainty computation approaches of UnCert-CoT.
\end{tcolorbox}

\section{Discussion}
\label{discussion}

\textbf{Baseline selection} The Structured Chain-of-Thought (SCoT) \cite{li2025structured} method fundamentally differs from our work in several key aspects. First, unlike the traditional CoT, the SCoT utilizes well-tailored prompts that include humane-curated structured reasoning instructions, providing additional organized information, our method and the self-planning generate outputs based on a simpler setup. Specifically, our method relies on a few-shot example composed of code and a natural language plan, without the specially designed prompts and expensive prompt engineering overhead that the SCoT required. 
Second, another fundamental difference lies in the generation paradigm of the decoding process. The SCoT employs a two-stage generation approach, wherein the LLM first produces a structured CoT. A reflection-based prompt engineering technique is then utilized to guide the LLM in verifying the structured CoT and mitigating potential noise. The refined structured CoT is subsequently used as input for the LLM to generate the corresponding code, which introduces significant computational overhead and implementation complexity. In contrast, UnCerT-CoT and all of the other baselines do not employs the reflection mechanism. Therefore, this study does not include a direct comparison between our UnCerT-CoT and the SCoT.

\begin{figure*}[tbp]
    \centering
    \includegraphics[width=0.98\linewidth]{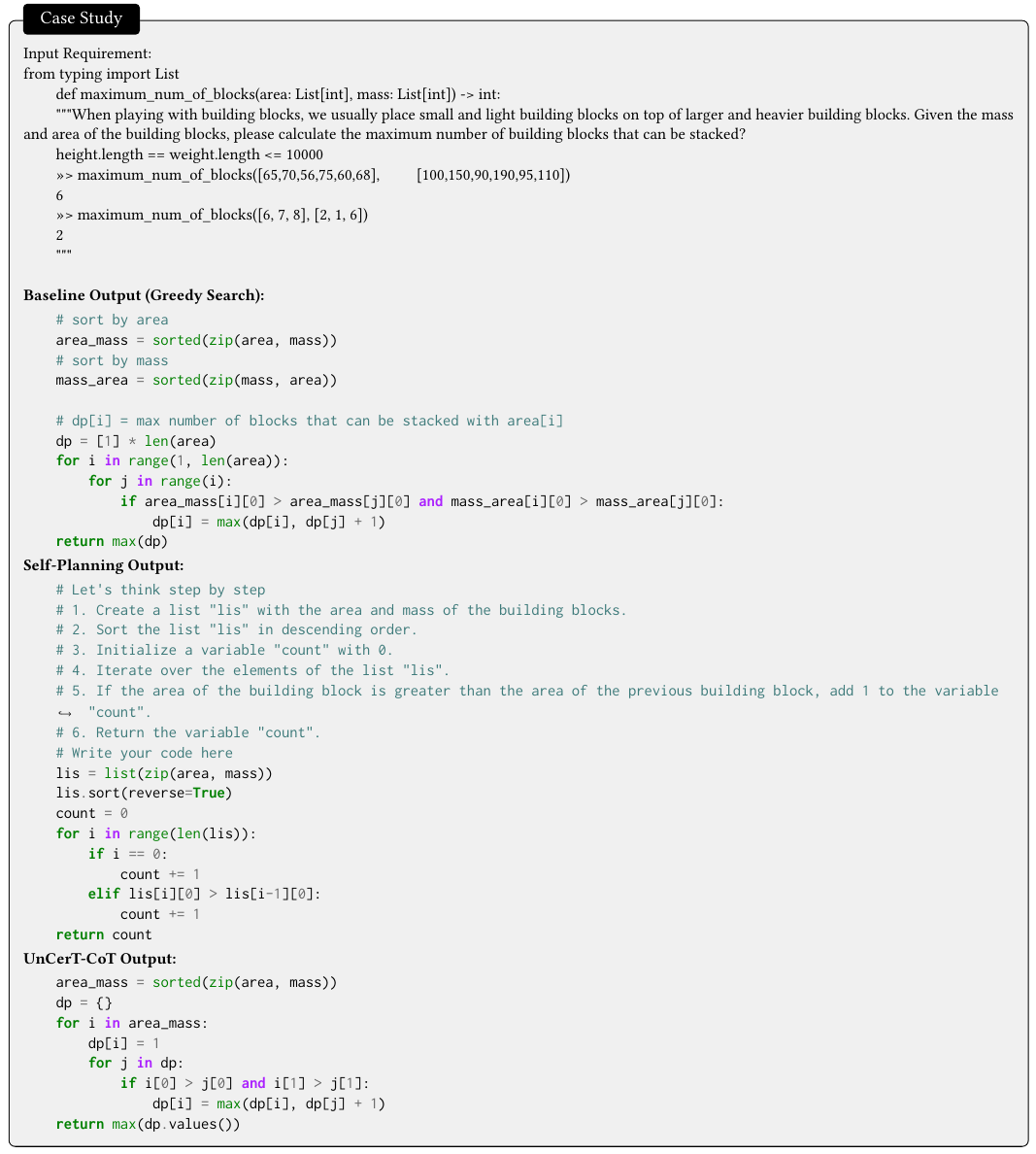}
    \caption{An example of the code generation results of UnCerT-CoT and state-of-the-art baselines.}
    \label{case_study}
\end{figure*}

\textbf{Case Study} We present a case study of a coding problem where the solution generated by our UnCerT-CoT method successfully passes all test cases, in contrast to the failure of both the base model and the self-planning method. The case is present in Figure \ref{case_study}.

\section{Related Work}

\subsection{Code Generation with LLMs}
The rapid development of LLMs, such as GPT-3 \cite{gpt-3} and GPT-4 \cite{GPT-4}, has significantly advanced the field of code generation. In recent years, to more precisely meet programming needs, researchers have developed Code LLMs specifically optimized for programming scenarios. These Code LLMs have demonstrated exceptional capabilities in various specific contexts, such as multi-language programming, code completion, debugging and fixing, and code refactoring \cite{codex,starcoder,codegeex}. Through training on large-scale codebases, Code LLMs can understand complex programming logic, syntactic structures, and developer intents, thereby significantly lowering the barrier to development and improving development efficiency \cite{FanGHLSYZ23,JiangLLMSurvey}. Currently, representative Code LLMs include CodeLlama \cite{codellama}, Deepseek-coder \cite{DeepSeek-Coder}, and Qwen-coder \cite{qwen}, which have been widely adopted in both open-source communities and the industry.
To advance the performance of LLMs in code generation tasks, software engineering research has introduced specialized enhancement techniques. These techniques focus on three key methodologies: First, prompting techniques to integrate software engineering expertise, such as domain knowledge and design patterns, to guide LLMs toward more complex solutions \cite{selfplanning,SCoT,ShrivastavaLT23}. Second, targeted fine-tuning optimizes LLMs using domain-specific datasets, sharpening their ability to generate contextually relevant code \cite{FineTuningCodeLLM,seed,Toolgen}. Finally, program-specific decoding leverages test cases, runtime feedback, and probabilistic sampling strategies to refine outputs, enhancing the accuracy and soundness of the model-generated code \cite{adapt,planning,MBR,monitor,debug}.

\subsection{Chain-of-Thought Prompting}


Chain-of-Thought (CoT) prompting was proposed by \citet{cot} to address the performance bottlenecks of LLMs in complex reasoning tasks such as mathematical problems and logical inference. Its core idea involves explicitly guiding models to generate intermediate reasoning steps that mimic human step-by-step problem-solving processes \cite{cot}. Subsequent research includes \citet{SuzgunSSGTCCLCZ23} who validated CoT's effectiveness in zero-shot scenarios, and \citet{KojimaGRMI22}, who demonstrated that the simple instruction "Let's think step by step" could activate LLMs' reasoning capabilities.
A notable derivative method for enhancing CoT is self-consistency \cite{selfconsistency}, which votes on the answers produced by multiple reasoning paths and selects the answer supported by the majority. Moreover, the latest advancements in LLMs, exemplified by OpenAI o1 \cite{opeai_o1} and Deepseek-R1 \cite{guo2025deepseek}, have integrated CoT capabilities as fundamental reasoning mechanisms through reinforcement learning. Recently, CoT has also been utilized in program generation. Self-planning \cite{selfplanning} approach leverages the CoT method for planning prior to code generation, thereby reducing the difficulty of problem-solving. Drawing inspiration from program structure, SCoT \cite{SCoT} incorporates sequential, branching, and looping structures into CoT with a reflection mechanism to enhance code generation.

\subsection{Uncertainty of LLMs}

As large language models (LLMs) continue to advance and find applications across various domains, understanding and quantifying uncertainty in their predictions has become increasingly essential. Uncertainty estimation provides valuable insights into the confidence of model outputs, which is particularly important for decision-making in high-stakes areas such as medical diagnosis \cite{fox1980evolution}\cite{simpkin2016tolerating}, where incorrect predictions can lead to severe consequences \cite{alkaissi2023artificial}. Additionally, uncertainty estimation plays a critical role in mitigating hallucinations in LLMs, offering a mechanism to identify when a LLM's answer falls outside its knowledge boundary \cite{li2023halueval}. Without effective uncertainty measures in transformer-based systems, it becomes challenging to rely on generated language as a trustworthy source of information \cite{kuhn2023semantic}. In this context, uncertainty encompasses the overall distribution of the model's outputs, reflecting the variability of predictions, whereas confidence refers to the likelihood of a specific prediction being correct. 

In \cite{codecalibration}, the authors investigate the relationship between the confidence levels expressed by LLMs in code tokens and the accuracy of those tokens with code completion tasks. Their findings reveal a strong calibration between the LLM’s entropy-based uncertainty and the correctness of generated code tokens across various LLMs. Specifically, the study demonstrates that when an LLM expresses high uncertainty in a particular token, there is a higher likelihood that the token is incorrect. 
This insight underscores the potential utility of uncertainty level in identifying and mitigating errors during code generation, as high-uncertainty predictions are more prone to inaccuracies. 

\section{Threats to Validity}
There are two main threats to the validity of our work.

\textbf{Threats to external validity} include the quality of the datasets. We select the datasets that contain various programming tasks ranging from easy to difficult. The problems are all human-curated and checked. We use two benchmarks including HumanEval \cite{codex}, and MHPP \cite{MHPP}.  
The coding problems in these benchmarks include handling variance in natural language inputs, understanding newly defined contexts, demonstrating commonsense, dealing with edge cases, following complex instructions, using mathematical and algorithmic knowledge, and showing familiarity with coding principles. This ensures the generalizability of our method on different types of programming tasks.
To verify the superiority of UnCerT-CoT, we consider three LLMs from different LLM families and with different sizes. (e.g. DeepSeek-Coder-6.7B -base\cite{DeepSeek-Coder}, CodeLlama-13B-python-hf \cite{codellama}, Qwen2.5-Coder-7B-Base\cite{Qwen-Coder}) as base models. We use PassRate as the evaluation metric. It is an execution-based metric that utilizes test cases to check the correctness of generated programs, ensuring the evaluation correctness of our results.

\textbf{Threats to internal validity} include the influence of the model hyper-parameter settings. 
The choice of hyperparameter values is crucial, and while we have selected configurations that provide positive results, different settings could potentially lead to variations in performance. In our experiments, we keep hyper-parameters ($k$, $m$, $t$) the same for all approaches. We experiment with different settings of $\tau$ to further test the upper bound of our method and illustrate the influence of changing the portion using the CoT method on the final results. Therefore, we believe that there is only a minor threat to the hyper-parameter choosing, and there might be room for further improvement.


\section{Conclusion}

In this paper, we propose UnCerT-CoT, a novel decoding method that improves code generation in LLMs by integrating uncertainty-aware mechanisms. UnCerT-CoT strategically applys the Chain-of-Thought method on points that require reasoning. Specifically, UnCert-CoT dynamically adapts CoT to generate intermediate reasoning steps based on the model’s uncertainty
during the code generation process. We utilize two confidence-based uncertainty measures: the Entropy-based and Probability Differential-based method. 
The experimental results show that the proposed UnCert-CoT substantially outperforms the state-of-the-art CoT methods with various LLMs. 
Moreover, the UnCert-CoT method demonstrates a certain level of robustness to hyperparameter settings.  

\newpage


\newpage
\bibliographystyle{ACM-Reference-Format}
\bibliography{reference}


\end{document}